\documentclass[final,5p,times,twocolumn]{elsarticle} 

\usepackage{amssymb}
\usepackage{amsmath}

\usepackage{siunitx}
\usepackage{subcaption}
\usepackage{graphicx}
\usepackage[colorlinks=true, linkcolor=blue, citecolor=blue, urlcolor=blue]{hyperref}
\usepackage[capitalize,nameinlink]{cleveref}
\crefname{figure}{Fig.}{Figs.}

\usepackage{lineno}

\begin{document}

\begin{frontmatter}



\title{Studies on the effect of low-fluence proton and neutron irradiation on n-type LGADs}

\author[label1,label2]{Veronika Kraus \corref{cor1}}
\author[label3]{Marcos Fernandez Garcia}
\author[label4]{Salvador Hidalgo} 
\author[label1]{Michael Moll}
\author[label4]{Jairo Villegas} 

\cortext[cor1]{Corresponding author. Email: \href{mailto:veronika.kraus@cern.ch}{veronika.kraus@cern.ch}}

\affiliation[label1]{organization={CERN, Organisation europénne pour la recherche nucléaire},
            addressline={Espl. des Particules 1},
            city={Genève},
            postcode={1217},
            country={Switzerland}}

\affiliation[label2]{organization={TU Wien, Faculty of Physics},
            addressline={Wiedner Hauptstraße 8-10},
            city={Vienna},
            postcode={1040},
            country={Austria}}

\affiliation[label3]{organization={Instituto de Física de Cantabria, IFCA (CSIC-UC)},
            addressline={Avda. los Castros},
            city={Santander},
            postcode={39005},
            country={Spain}}

\affiliation[label4]{organization={Instituto de Microelectrónica de Barcelona (IMB-CNM-CSIC)},
            addressline={Cerdanyola del Vallès},
            city={Barcelona},
            postcode={08193},
            country={Spain}}

\begin{abstract}
The presented study investigates the effects of low fluences from \SI{5e12}{} up to \SI{1e14}{\text{particles}/\centi\meter\squared} 
of \SI{60}{\mega\electronvolt} proton and neutron irradiation on n-type Low Gain Avalanche Detectors (nLGADs). An nLGAD is a silicon sensor with a highly doped gain layer that enables controlled charge multiplication via impact ionization. In contrast to the well-established p-type LGADs for high-energy physics (HEP) applications, nLGADs are optimized for the detection of low-penetrating particles such as UV photons and soft X-rays. In addition to studying their potential application in environments with radiation backgrounds, these novel devices also enable the exploration of the underlying phenomenology arising from the combination of n-type bulk material with a gain layer, which degradation was previously studied predominantly in the context of p-type LGADs. The irradiation effects were characterized through measurements of the leakage current and capacitance with increasing bias voltage (I-V and C-V), revealing systematic and fluence-dependent behavior related to space charge sign inversion (SCSI) of the n-type bulk material, which especially alters the electric field in the sensor and thus the depletion behavior. Additionally, annealing studies were performed to assess both beneficial and reverse annealing regimes with isothermal and isochronal annealing. The findings are consistent with previous high-energy proton studies and contribute to a deeper understanding of the fundamental behavior of nLGADs under irradiation. 
\end{abstract}


\end{frontmatter}

\section{Introduction}
\label{intro}
Low Gain Avalanche Detectors (LGADs) are a class of silicon sensors designed to provide moderate internal signal amplification through controlled charge multiplication. They feature a highly doped gain layer which creates a high electric field region where charge carriers undergo impact ionization, leading to amplification of the primary signal charge. LGADs implemented as $n^{++}-p^{+}-p$, where $p^+$ refers to the gain layer, have shown to be promising timing detectors for Minimum Ionizing Particles (MIPs) and will therefore be implemented in the high-luminosity detector upgrade of the Large Hadron Collider (HL-LHC) as precision timing detectors to cope with high occupancy in future tracking. However, the p-type LGAD, which has recently become well-established in the field of high-energy physics (HEP), is not optimized for the detection of low penetrating particles. To address this limitation, n-type LGADs (nLGADs) have been developed. These sensors invert the doping profile, featuring an n-type bulk with an $n^+$ implant as gain layer. Internal gain in silicon sensors, also in devices with n-type bulk, has previously been realized in Avalanche Photodiodes (APDs), as for example described in \cite{Musienko2000}. nLGADs apply this principle in a controlled and application-specific manner, particularly optimized for precise timing and the detection of low-penetrating particles. This opens up the possibility of using LGAD technology in areas beyond HEP and exploiting the advantage of internal signal gain for the detection of low-penetrating protons, soft X-rays and UV photons. 
The present work describes studies on the effects of irradiation damage induced by different particle types in the novel nLGAD concept. Beyond practical considerations, the irradiation-induced degradation of nLGADs provides a valuable opportunity to probe fundamental material properties and explore the underlying phenomenology, particularly in comparison to traditional p-type LGADs. Previous studies have indicated that nLGADs are less radiation hard than their p-type counterparts \cite{Kraus2025}, which motivates dedicated investigations at low fluences from \SI{5e12}{} up to \SI{1e14}{\text{particles}/\centi\meter\squared}. Such insights can as well be of relevance to current developments in HEP, such as the compensated LGAD concept \cite{Sola2024}, and for suggested nLGAD applications in environments with high particle backgrounds, for example in space or nuclear fusion experiments \cite{Han2022}.

\section{Materials and methods}

\begin{figure*}[t]
    \centering
    \begin{minipage}{0.5\textwidth}
        \centering
        \includegraphics[width=0.7\textwidth]{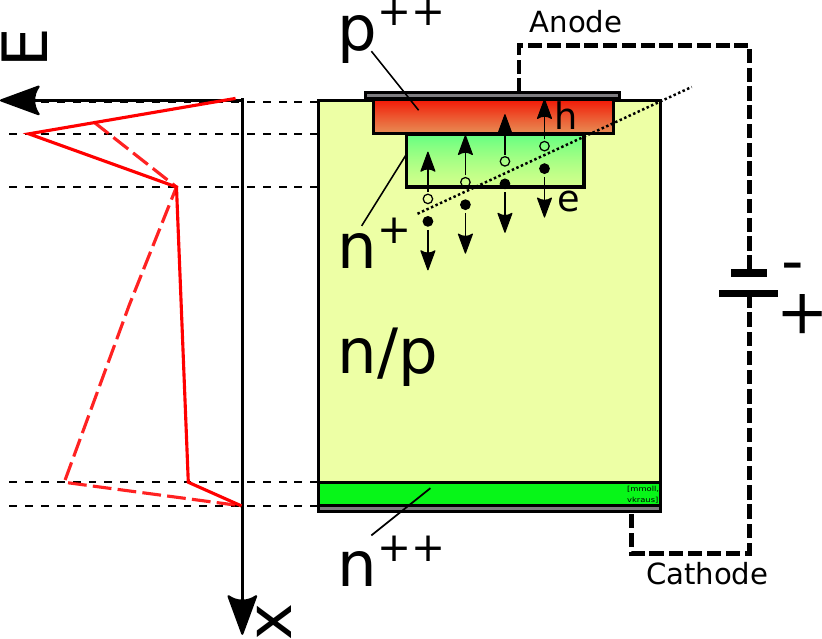}
        \par\vspace{0.5ex}
        (a)
    \end{minipage}%
    \begin{minipage}{0.5\textwidth}
        \centering
        \includegraphics[width=0.5\textwidth, angle=180]{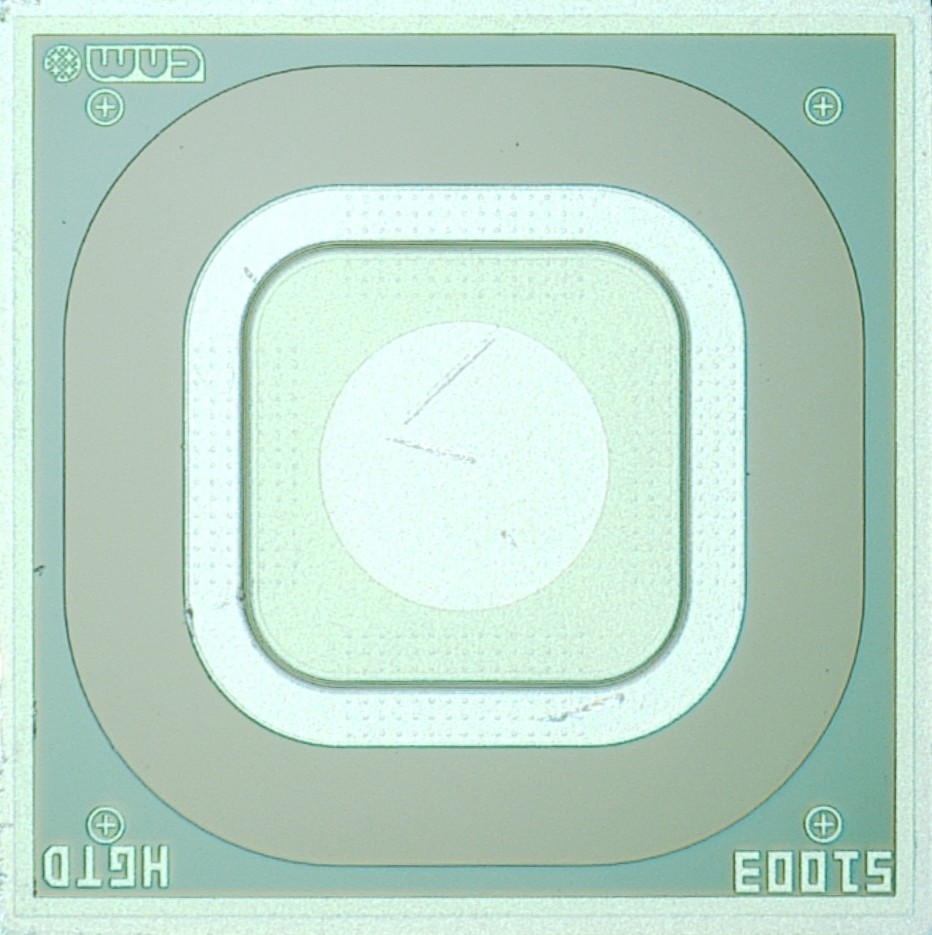}
        \par\vspace{0.5ex}
        (b)
    \end{minipage}
    \caption{a: Schematic cross-section of the nLGAD detector structure. b: Top-view microscope image of one sample.}
    \label{fig:nLGAD_overview}
\end{figure*}

\begin{table*}[t]
    \centering
    \begin{tabular}{|c|c|c|c|}
    \hline
    \multicolumn{1}{|l|}{\textbf{Number of devices}} &
      \textbf{Particle type} &
      \multicolumn{1}{l|}{\textbf{Fluences {[}particles/cm$^2${]}}} &
      \multicolumn{1}{l|}{\textbf{n$_{\text{eq}}$ Fluences}} \\ \hline
    5 nLGADs + 5 PiNs & neutrons       & \begin{tabular}[c]{@{}c@{}}\SI{5e12}{}\\ \SI{7e12}{}\\ \SI{2e13}{}\\ \SI{4e13}{}\\ \SI{6e13}{}\end{tabular} & -                                                      \\ \hline
    4 nLGADs + 4 PiNs &
      60\,MeV protons &
      \begin{tabular}[c]{@{}c@{}}\SI{5e12}{}\\ \SI{1e13}{}\\ \SI{5e13}{}\\ \SI{1e14}{}\end{tabular} &
      \begin{tabular}[c]{@{}c@{}}\SI{8.0e12}{}\\ \SI{1.6e13}{}\\ \SI{8.0e13}{}\\ \SI{1.6e14}{}\end{tabular} \\ \hline
    \end{tabular}
    \caption{Overview of nLGADs and reference PiNs included in this study with corresponding irradiation type (neutrons or \SI{60}{MeV} protons) and fluences.}
    \label{tab:all_fluences}
\end{table*}

This section provides an overview of the tested devices and the irradiation campaigns. The nLGADs and corresponding reference PiNs were exposed to neutrons and \SI{60}{MeV} protons to probe the response of the devices to different particle types and fluences.

\subsection{Devices under test}
The devices under test are produced by IMB-CNM and are implemented as $p^{++}-n^{+}-n$, as can be seen from the sketch of the cross section in \autoref{fig:nLGAD_overview}. This layout offers an advantage for the detection of low-penetrating particles, since electrons produced in the first few \SI{}{\micro\m} of the structure initiate impact ionization in the gain layer with a higher rate compared to holes. More details on the working principle of nLGADs can be found in \cite{Villegas2023}. The first nLGAD production run R16375 from IMB-CNM is fabricated on n-type high resistivity FZ wafers with an active thickness of \SI{275}{\micro\meter}. Included in this study are single test devices with an active area of \SI{1}{\milli\meter} $\times$ \SI{1}{\milli\meter} and full-cover metallisation on top. Reference PiN diodes from the same wafer without a gain layer are also available.

\subsection{Irradiation}
The available 18 devices were irradiated with different types of particles and particle energies. All tested fluences in this study are relatively low compared to the typical levels in HEP applications, where conventional p-type LGADs in the HL-LHC upgrades of CMS and especially ATLAS are expected to face up to \SI{2.5e15}{\text{n\textsubscript{eq}}/\centi\meter\squared} end-of-lifetime fluences. Tested fluences in this study reach up to a maximum of \SI{1e14}{\text{particles}/\centi\meter\squared} (\SI{1.6e14}{\text{n\textsubscript{eq}}/\centi\meter\squared}). This is because previous studies have indicated that nLGADs are not as radiation hard due to space charge sign inversion (SCSI) and a presumably faster degeneration of the gain layer \cite{Kraus2025}, which motivates the investigation at lower fluence regimes to study the onset of degradation in more detail.
Five nLGAD samples and corresponding reference PiNs were irradiated with neutrons at the TRIGA nuclear reactor of the Jo\v{z}ef Stefan Institute (JSI) in Ljubjana. The targeted fluences, listed in \autoref{tab:all_fluences}, range from \SI{5e12}{\text{neutrons}/\centi\meter\squared} to \SI{6e13}{\text{neurons}/\centi\meter\squared} and were achieved by the irradiation facility within an accuracy of 10\%.

\begin{figure*}[t]
    \centering
    \begin{subfigure}{0.45\textwidth}
        \centering
        \includegraphics[width=\textwidth]{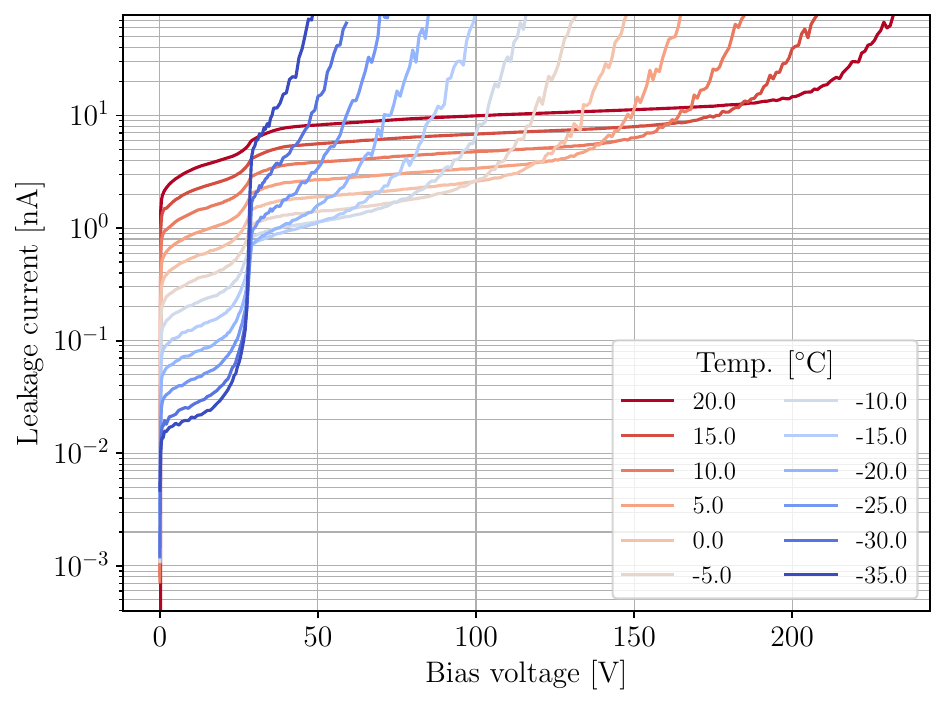}  
        \caption{}
        \label{subfig:IV_T}
    \end{subfigure}
    \hfill
    \begin{subfigure}{0.45\textwidth}
        \centering
        \includegraphics[width=\textwidth]{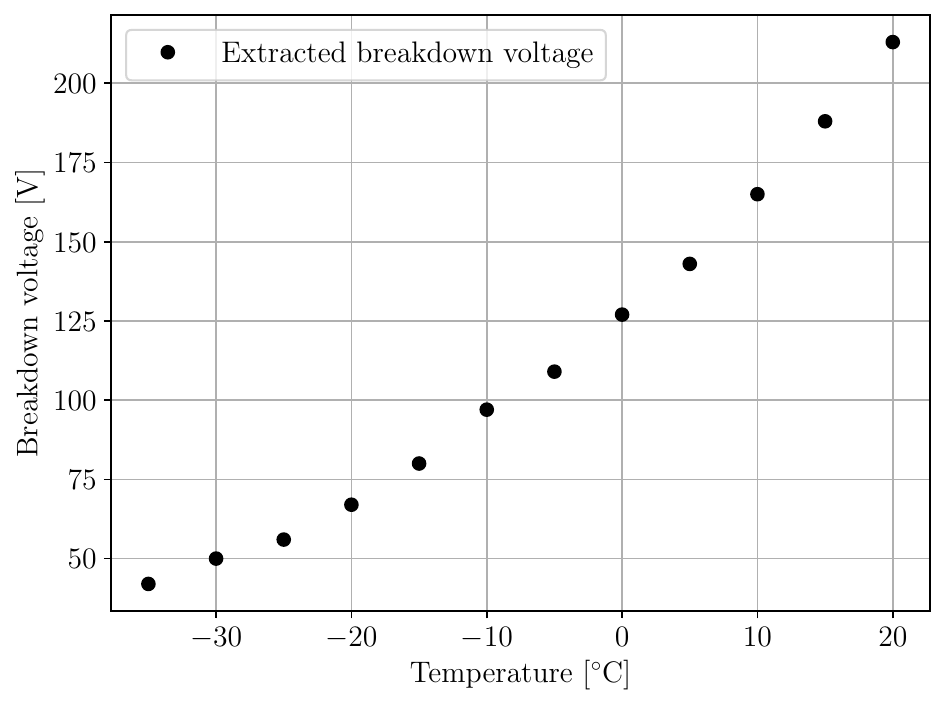}  
        \caption{}
        \label{subfig:Vbias_T}
    \end{subfigure}
    \hfill
    \begin{subfigure}{0.45\textwidth}
        \centering
        \includegraphics[width=\textwidth]{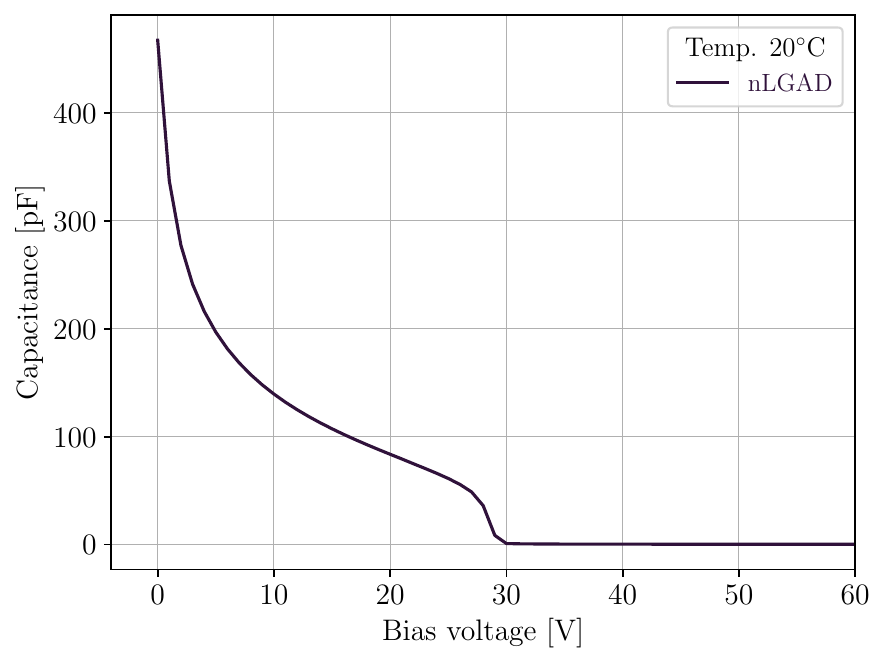} 
        \caption{}
        \label{subfig:CV_non_irrad}
    \end{subfigure}
    \caption{a: I-V characteristic of an nLGAD before irradiation measured at various temperatures. b: From I-V curves extracted breakdown voltages plotted as a function of measurement temperature. A shift of breakdown towards higher bias voltages with rising temperature can be observed, coming from the temperature dependence of impact ionization in the gain layer. c: }
    \label{fig:Impact_Ion_T}
\end{figure*}

Furthermore, four nLGAD samples and corresponding reference PiNs were irradiated with \SI{60}{\mega\electronvolt} protons at the AIC-144 cyclotron of the Institute of Nuclear Physics of the Polish Academy of Sciences (IFJ PAN) in Krakow. Again, the individual fluences are listed in \autoref{tab:all_fluences}, with an accuracy well below 5\%, as the facility was originally established for ocular cancer therapy and has been optimized for delivering highly precise irradiation doses. In order to compare damage induced from the two different particle types in terms of their Non-Ionizing Energy Loss (NIEL), the hardness factor $\kappa$ is used to convert the proton fluences to \SI{1}{MeV} neutron equivalent fluences ($\mathrm{n_{eq}}$). Since there is no specified $\kappa$ given by the facility, a value was determined from measuring the increasing leakage currents of the irradiated PiNs, shown in \autoref{subfig:IV_PiN}. The leakage current values for increasing fluences are fitted after full depletion of the PiNs, once the current reaches a stable plateau over a broad voltage range. The relation between leakage current and $\kappa$ is given by the following equations:  
\begin{equation}
\frac{\Delta I}{V} = \alpha \cdot \phi_p \quad \text{and} \quad \kappa = \frac{\alpha}{\alpha_{\mathrm{n_{eq}}}} , 
\end{equation}
with $V$ denoting the active sensor volume, $\phi_p$ the proton fluence, and $\alpha$ the current-related damage rate extracted from the data. $\alpha_{\mathrm{n_{eq}}}$ is a reference value for the current-related damage rate corresponding to \SI{1}{MeV} neutron equivalent fluence, taken from \cite{Moll1999}. The resulting hardness factor is determined to be $\kappa = 1.60 \pm 0.03$ which is in line with theoretical values given by Huhtinen and Summers \cite{HuhtinenSummers2000}.

\section{Electrical characterization before irradiation}

\label{El. char irrad}
\begin{figure*}[t]
    \centering
    \begin{subfigure}{0.45\textwidth}
        \centering
        \includegraphics[width=\textwidth]{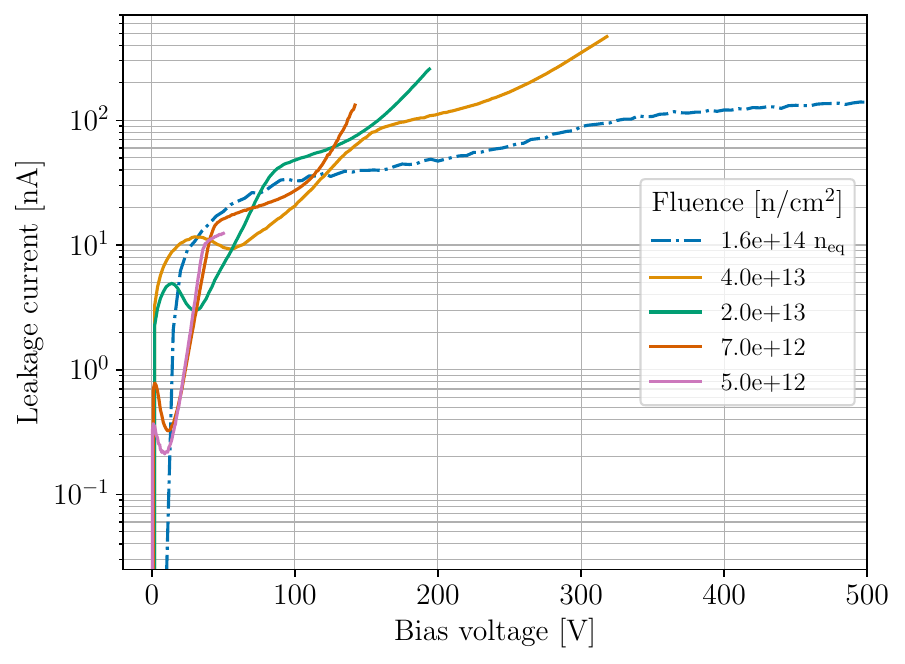}
        \caption{}
        \label{subfig:IV_irrad}
    \end{subfigure}
    \hfill
    \begin{subfigure}{0.45\textwidth}
        \centering
        \includegraphics[width=\textwidth]{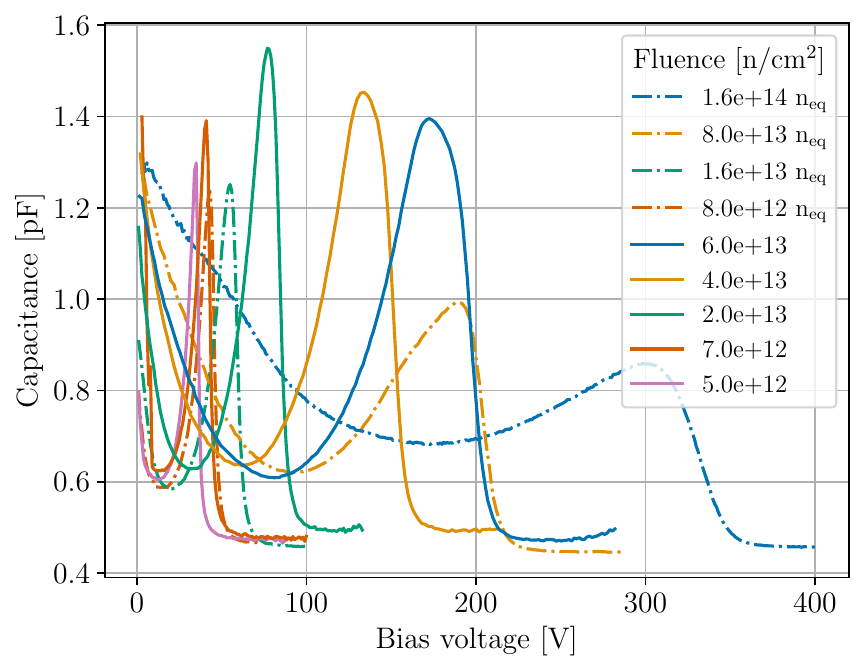} 
        \caption{}
        \label{subfig:CV_Irrad}
    \end{subfigure}
    \caption{Electrical characterization of nLGADs after irradiation performed at \SI{-20}{\degreeCelsius}. a: I-V characteristics shown for representative fluences. b: C-V characteristics at \SI{1}{kHz}.}
    \label{fig:IV_CV}
\end{figure*}

\begin{figure*}[t]
    \centering
    \begin{subfigure}{0.45\textwidth}
        \centering
        \includegraphics[width=\textwidth]{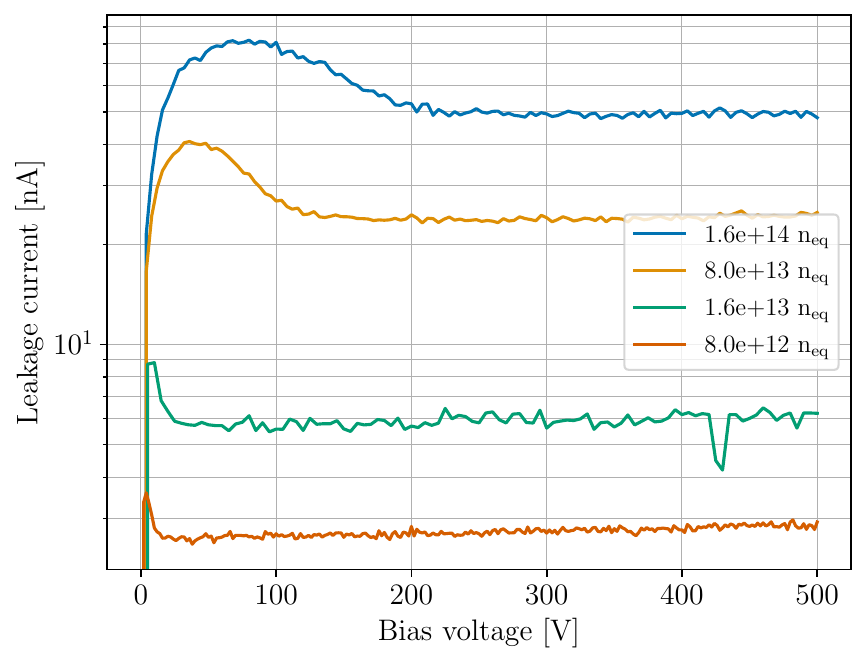} 
        \caption{}
        \label{subfig:IV_PiN}
    \end{subfigure}
    \hfill
    \begin{subfigure}{0.45\textwidth}
        \centering
        \includegraphics[width=\textwidth]{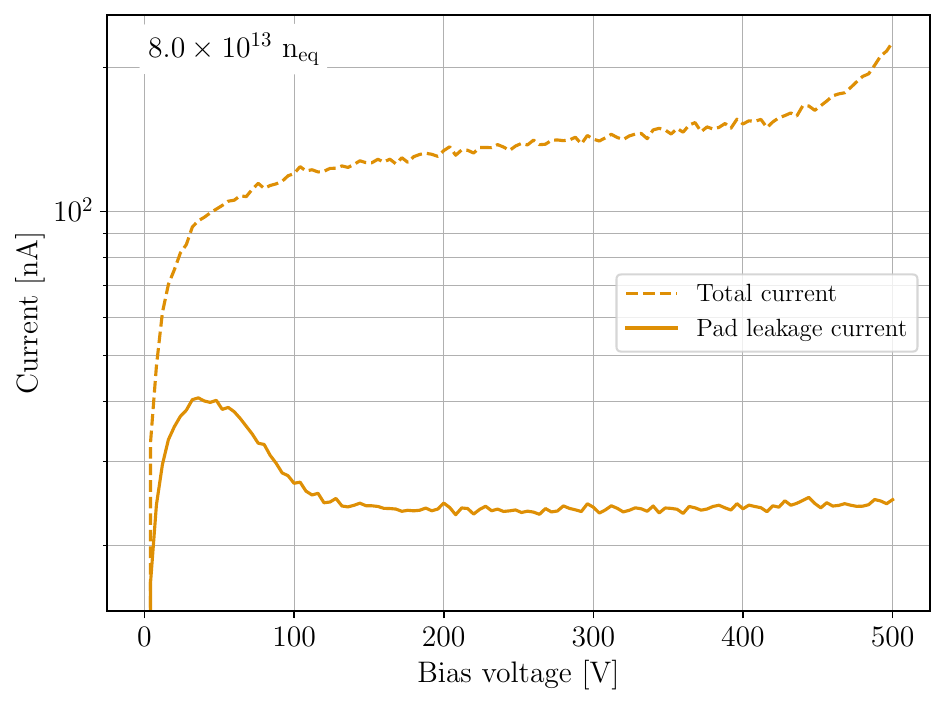} 
        \caption{}
        \label{subfig:Pad_vs_total_current}
    \end{subfigure}
    \caption{a: I-V characteristics of the proton irradiated PiNs. At low bias voltages, the leakage current increases with the expanding depletion region, but decreases once the guard ring becomes active and limits the contributing area to the pad. This behavior is confirmed in b: Comparison between total and pad current in the device. The total current continuously increases with bias voltage, unaffected by the guard ring.}
    \label{fig:PiN}
\end{figure*}

For the electrical characterization the probe-station at the SSD laboratory at CERN was used. The samples are placed directly on a chuck that can be temperature controlled and was set to \SI{20}{\degreeCelsius} for the described measurements before irradiation. The tested sample is contacted via two micro-positioning needles, one on the pad for signal readout and one on the guard ring, with the guard ring always grounded. A high-voltage bias is applied to the sample backside via the chuck during electrical characterization. Measurements of the leakage current and capacitance with increasing bias voltage (I-V and C-V) reveal gain layer depletion around \SI{28}{\volt} and breakdown between \SI{200}{\volt} and \SI{250}{\volt} at \SI{20}{\degreeCelsius}, as can be seen in \autoref{fig:Impact_Ion_T}.

\subsection{Temperature dependence of impact ionization}
Before irradiation, I-V characteristics for one nLGAD were measured for various temperatures between \SI{-35}{\degreeCelsius} and \SI{+20}{\degreeCelsius}. A temperature dependence of the breakdown voltage is clearly visible in \autoref{subfig:IV_T}, with higher temperatures leading to a later onset of breakdown with regards to the applied bias voltage. The temperature dependent increase in breakdown voltage is shown in \autoref{subfig:Vbias_T}. The breakdown voltage was extracted from the I-V characteristics by evaluating $\frac{dI}{dV} \cdot \frac{V}{I}$ which enhances the steep slope near breakdown. The breakdown voltage was defined as the first point exceeding a chosen threshold within a voltage range above \SI{30}{V}. The breakdown voltage shifts within the tested temperature range from below \SI{50}{V} at low temperature to above \SI{220}{V}. This behavior can be attributed to the strong temperature dependence of impact ionization, leading to the multiplication of charge carriers in the gain layer. At higher temperatures, increased phonon scattering reduces the carrier mean free path, which lowers the probability of impact ionization and thus increases the breakdown voltage. The extracted breakdown voltages fitted with a linear model give a temperature coefficient of $\approx 3\,\text{V/K}$.

\section{Electrical characterization after irradiation}

The electrical characterization after irradiation was performed at \SI{-20}{\degreeCelsius} to reduce the elevated leakage current in irradiated silicon. Furthermore, all nLGADs and reference PiNs were subjected to a first annealing step of \SI{4}{min} at \SI{80}{\degreeCelsius} before the presented I-V and C-V characterization in order to balance out any short term annealing at room temperature during irradiation and handling of the samples. 

\subsection{nLGAD I-V characteristics}
Similar effects on the nLGAD I–V characteristics can be seen after \SI{60}{MeV} proton and neutron irradiation. Representative curves are shown in \autoref{subfig:IV_irrad}. The three main effects observed after irradiation, also consistent with findings from previous \SI{23}{GeV} proton irradiation of nLGADs \cite{Kraus2025}, are:

\begin{enumerate}
\item An increase in leakage current, as expected due to radiation induced defect generation in the bulk.
\item The steep increase in leakage current associated with gain layer depletion becomes less pronounced and shifts towards higher bias voltages with increasing fluence.
\item A characteristic "dip" in the leakage current is observed before the sensor reaches full depletion. This effect is attributed to the guard ring and also depends on the irradiation fluence, becoming broader and less distinct with higher particle exposure. The formation of this "dip" is described in more detail in \autoref{Characterization of reference PiN structures} for reference PiNs, as the effect is easier to detach from other contributions in the simpler structures.
\item Furthermore, the breakdown voltage increases with irradiation, indicating a reduced electric field strength in the gain layer when it degrades with increasing fluences.
\end{enumerate}

The macroscopic changes of nLGAD characteristics with irradiation differ from those observed in traditional p-type LGADs, where increasing fluences lead to a shift of gain layer depletion towards lower applied bias voltages as the gain layer becomes ineffective \cite{Curras2023}, \cite{Curras_Rivera2023}. This differences originate from the space charge sign inversion (SCSI) of the n-type silicon bulk, a well-known phenomenon in n-type PiN diodes that arises from the combined effects of donor removal and the introduction of irradiation-induced acceptor-like defects as reported e.g., in \cite{Pitzl1992}, \cite{Matheson1996}. This behavior has been widely studied in high-resistivity n-type silicon sensors subjected to irradiation with various particles, particularly protons and neutrons, within a fluence range comparable to the present study. The above-mentioned sources report that PiN silicon detectors undergo space charge sign inversion after irradiation at fluences on the order of \SI{1e13}{\text{particles}/\centi\meter\squared}.

While p-type LGADs typically retain a top-to-backside depletion, where the gain layer depletes first after irradiation, SCSI leads to a reversal of the effective space charge in the bulk of n-type structures with gain layer, resulting in a modified electric field configuration and depletion starting from the backside. As demonstrated in a previous publication \cite{Kraus2025}, measuring depth profiles of the electric field in an irradiated nLGAD after SCSI, a second junction appears at the backside of the device at the boundary to the strongly doped $n++$ contact. This is reflected in a shift of the steep rise in the I-V curve to higher rather than lower bias voltages with increasing particle fluence, whereby gain layer depletion voltage no longer corresponds to this steep rise. Therefore, the study of irradiation-induced degradation particularly of the nLGAD gain layer remains challenging due to the complex electric field structures that develop after SCSI.

\subsection{Characteristics of reference PiN structures}
\label{Characterization of reference PiN structures}
The reference PiN structures have the same layout as the nLGADs, but without the additional gain layer. Therefore, their n-type bulk is affected by SCSI, and irradiation induced changes can be observed. I-V curves of the present CNM PiN structures measured after low-fluence proton irradiation are shown in \autoref{subfig:IV_PiN}. After irradiation to a certain fluence, the leakage current initially increases with bias voltage, then decreases, and eventually levels off at higher voltages. At low voltages, the leakage current increases with applied bias voltage as the depleted volume grows. This behavior is attributed, as for the nLGADs, to the change in depletion behavior after irradiation, where the bulk starts to deplete from the backside. Once the depletion region reaches the front side and the guard ring starts to pinch in, the current decreases, as only the active area inside the guard ring contributes to the measured current. The observed behavior enhances with increasing irradiation fluence, indicating a dependence on radiation damage. No breakdown is observed within the scanned voltage range, as the measurement setup is limited to a maximum of \SI{500}{V}. Similar effects are observed after neutron irradiation. In order to confirm that the shape of the I–V curve arises from the contacted guard ring, \autoref{subfig:Pad_vs_total_current} shows the leakage current measured at the pad, compared to the total device current, which is unaffected by the guard ring. The pad current is measured using a Picoammeter, which is typically used for standard leakage current measurements, while the total current is read from the Sourcemeter that also supplies the bias voltage.

\subsection{nLGAD C-V characteristics}

\begin{figure}[t]
\centering
\includegraphics[scale=0.48]{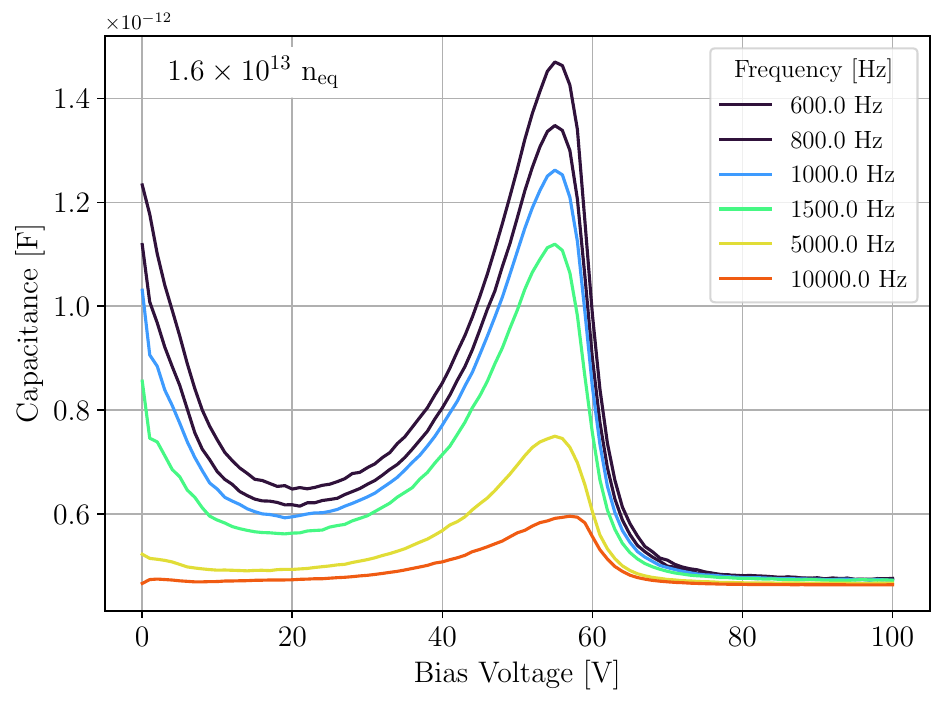}
\caption{Exemplary C-V curves of one proton irradiated nLGAD ($1.6\times10^{13}\ \mathrm{n_{eq}}$) for varying frequencies from \SI{600}{Hz} to \SI{10}{kHz}. A clear frequency dependence of the peak at full depletion, arising from resonances with defect charges, can be observed.}
\label{fig1:CV_frequ_dependent}
\end{figure}

C-V characteristics for neutron and proton irradiated samples were measured at a capacitance of \SI{1}{kHz}. The measured curves are presented in \autoref{subfig:CV_Irrad}. Before the capacitance reaches a constant, minimal value, indicating that the sensor is fully depleted, a peak can be observed in the C-V characteristics of all samples, which shifts to higher bias voltages and widens with increasing particle fluence. The peaks are higher and narrower after neutron irradiation compared to broader peaks after proton irradiation. The peak arising in the C-V characteristics for both proton and neutron irradiated samples originates from resonances with defect charges between two fields touching, once the nLGAD reaches full depletion. As already evident from the I-V characteristics, the electric field with applied bias voltage starts building from the backside of the device. A second field also starts building from the gain layer at the topside of the device. Both depletion regions extend until they merge in the bulk at full depletion, leading to the formation of the observed peak. The resonance is dependent on the frequency at which the measurement of the capacitance gets performed. The frequency dependence, tested from \SI{600}{Hz} to \SI{10}{kHz} is shown exemplary for a \SI{60}{MeV} proton irradiated sample in \autoref{fig1:CV_frequ_dependent}. The same trend is observed in all tested nLGADs after irradiation, revealing a distinct phenomenological behavior, which is in good agreement with the established understanding of the creation of irradiation-induced defects in silicon detectors.

\section{Annealing studies}

\begin{figure}[t]
\centering
\includegraphics[scale=0.48]{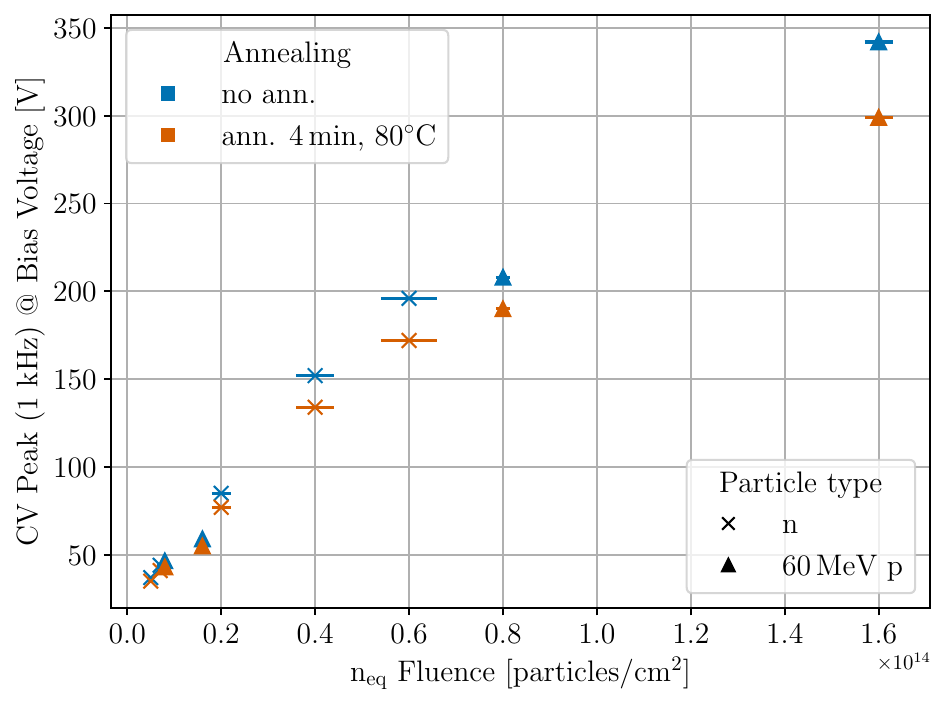}
\caption{Effect of annealing (\SI{4}{min} at \SI{80}{\degreeCelsius}) shown in terms of the shift in full sensor depletion, represented by the voltage at which the C-V curve exhibits a resonance peak, for all neutron and proton irradiated nLGADs. The observed shift of full depletion to lower bias voltages indicates that the first annealing step lies in the beneficial regime.}
\label{fig:1st_step_4min_80C}
\end{figure}

Post-irradiation annealing is a widely used method to restore silicon sensor properties and has been systematically studied for n-type PiN detectors in previous work, e.g., \cite{Moll1999} or \cite{Ziock1993}. In this part of the paper, the effects of annealing on nLGADs shall be investigated and related to known effects in n-type material in general. As mentioned in \autoref{El. char irrad}, a first annealing step of \SI{4}{min} at \SI{80}{\degreeCelsius} was performed for all samples in order to balance out any short annealing times that may occur and ensure that all samples are in a comparable condition. C-V measurements at a frequency of \SI{1}{kHz} were also carried out before the first annealing step. The peak in the C-V curve, which occurs when the sensor reaches full depletion, was chosen as the parameter used for comparison, since it is a good measure of changing electric fields and bulk properties with irradiation. \autoref{fig:1st_step_4min_80C} shows a comparison of the extracted peak positions from the C-V curves as a function of fluence, both before and after annealing, for all samples irradiated with neutrons and protons. It can be seen, that the peak position and therefore full sensor depletion can be shifted towards lower applied bias voltages with this first annealing step. In addition to the effect of beneficial annealing shifting full depletion to lower bias voltages for all samples, the effect is more pronounced for higher fluences.

\subsection{Isochronal annealing}

\begin{figure}[t]
\centering
\includegraphics[scale=0.48]{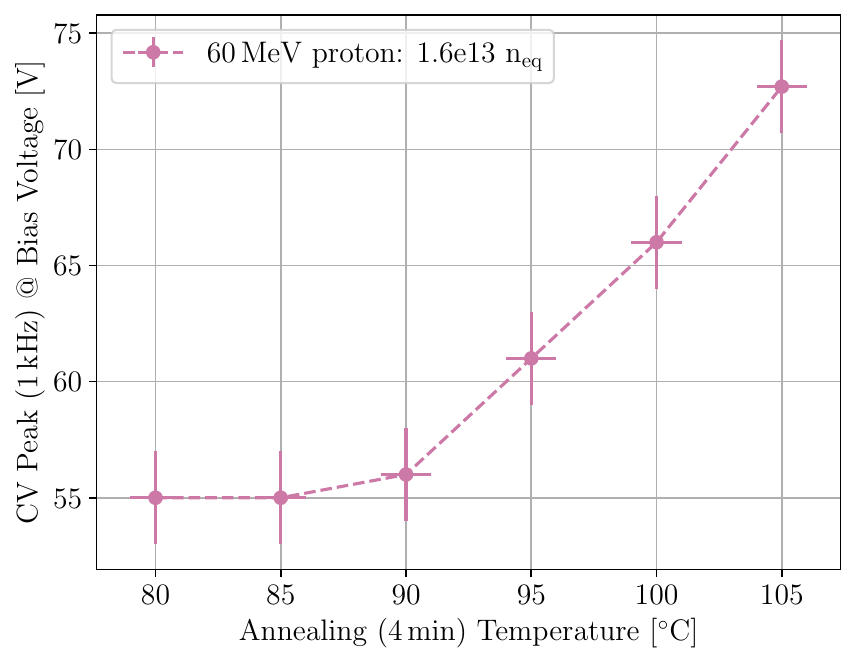}
\caption{Isochronal annealing of one proton irradiated nLGAD for successively higher temperatures at equal time intervals. The chosen measurement range shows the onset of the regime of reverse annealing.}
\label{fig:Annealing_isochron}
\end{figure}

For isochronal annealing, one proton irradiated nLGAD (fluence \SI{1e13}{particles/cm^2} or \SI{1.6e13}{n_{eq}}) was annealed at successively higher temperatures for equal time intervals, allowing the study of changes in the bulk as a function of temperature. Temperatures ranging from the first annealing step at \SI{80}{\degreeCelsius} up to \SI{105}{\degreeCelsius} in \SI{5}{\degreeCelsius} steps were tested, with an annealing time of \SI{4}{min} for each step. C-V characteristics after each annealing step were measured at \SI{1}{kHz}. Again, the peak in the C-V curve occurring when the sensor reaches full depletion is extracted from each measurement curve. \autoref{fig:Annealing_isochron} shows the peak position plotted against the increasing annealing temperatures. The plot shows that sensor full depletion is shifting to increasingly higher bias voltages from the initial \SI{50}{V} to over \SI{70}{V}. A region of beneficial annealing is not visible and most likely bypassed. 

\subsection{Isothermal annealing}

\begin{figure*}[t]
    \centering
    \begin{subfigure}{0.45\textwidth}
        \centering
        \includegraphics[width=\textwidth]{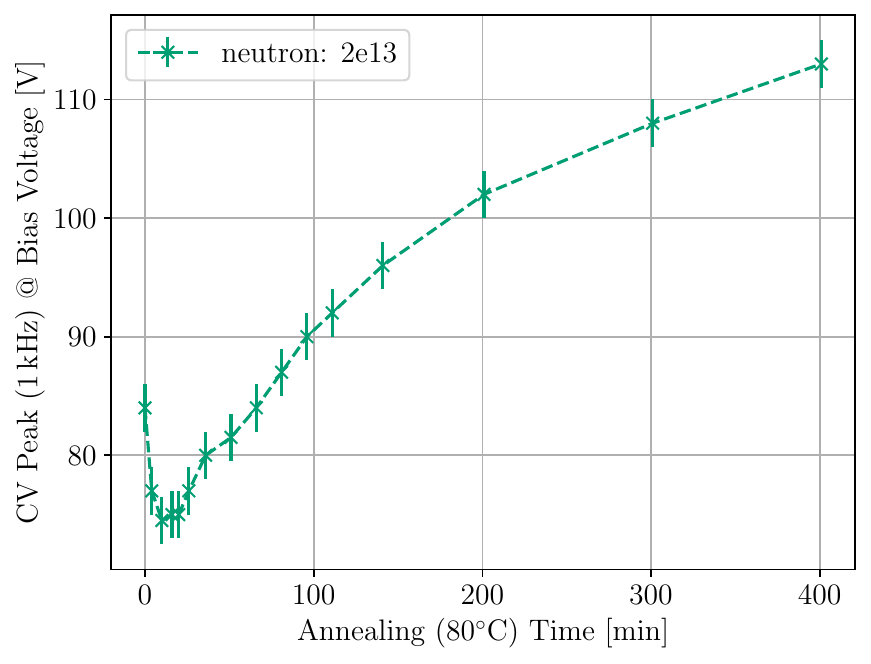} 
        \caption{}
        \label{subfig:Isotherm_n}
    \end{subfigure}
    \hfill
    \begin{subfigure}{0.45\textwidth}
        \centering
        \includegraphics[width=\textwidth]{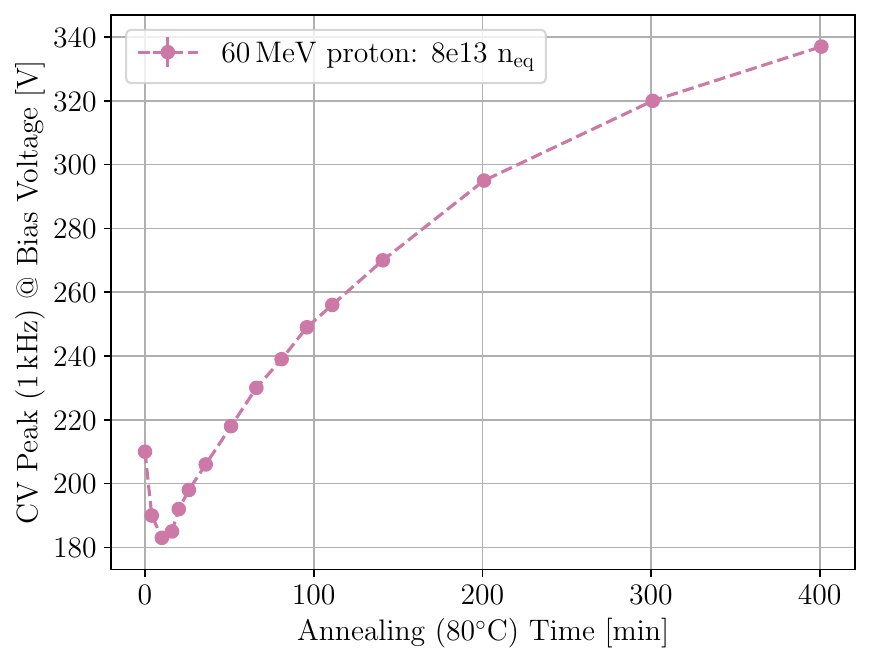} 
        \caption{}
        \label{subfig:Isotherm_p}
    \end{subfigure}
    \vskip\baselineskip
    \begin{subfigure}{0.45\textwidth}
        \centering
        \includegraphics[width=\textwidth]{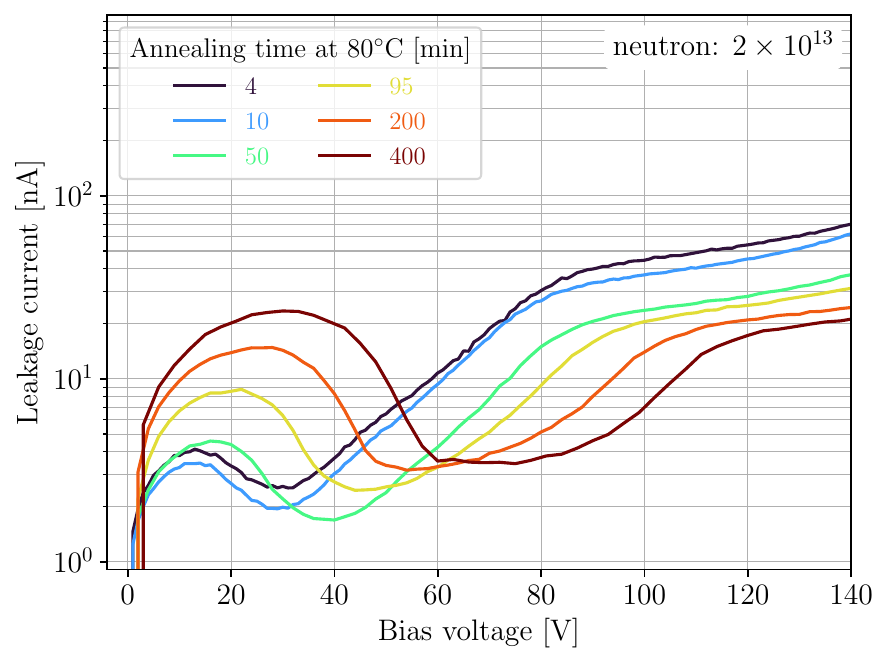} 
        \caption{}
        \label{subfig:Change_IV}
    \end{subfigure}
    \hfill
    \begin{subfigure}{0.45\textwidth}
        \centering
        \includegraphics[width=\textwidth]{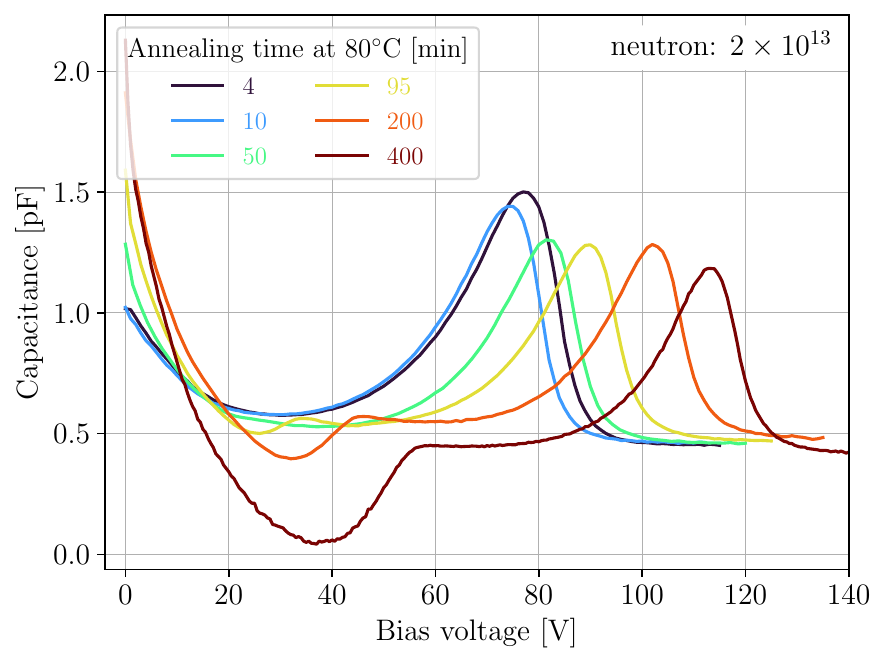} 
        \caption{}
        \label{subfig:Change_CV}
    \end{subfigure}
    \caption{Results after isothermal annealing. a and b: Effect of annealing for neutron and proton irradiated nLGAD at \SI{80}{\degreeCelsius} shown in terms of the shift in full sensor depletion with increasing time steps, represented by the voltage at which the C-V curve exhibits a resonance peak. Beneficial annealing after $\approx$ \SI{10}{min}. c and d: I-V and C-V characteristics for increasing annealing times, exemplary shown for the neutron irradiated sample. Same effects can be observed after proton irradiation.}
    \label{fig:Annealing}
\end{figure*}

The temperature dependence of mechanisms leading to defect annealing is given by an Arrhenius relation 
\begin{equation}
    R(T) \propto \exp\left(-\frac{E_a}{k T}\right)
\end{equation}
with $E_a$ being the activation energy \cite{Moll1999}. It is evident that higher temperatures significantly increase the rate of annealing. To complement the faster isochronal annealing, which skipped over the beneficial annealing range, two additional samples were annealed isothermally. One neutron irradiated sample with a fluence of \SI{2e13}{n/cm^2} was chosen, and one proton irradiated with a fluence of \SI{8e13}{n_{eq}}. Isothermal annealing was performed at \SI{80}{\celsius}, using progressively increasing annealing times from \SI{4}{min} up to \SI{400}{min}. The results presented in \autoref{subfig:Change_IV} and \autoref{subfig:Change_CV} show how the I-V and C-V characteristics change with increasing annealing time. The observed "dip" in the I-V curves becomes stronger pronounced, while a valley forms at the corresponding bias voltages in the C-V curves, which shows the significant changes in the electric fields with annealing. 

The evolution of the C-V peak voltage during isothermal annealing at \SI{80}{\degreeCelsius} for neutron and proton irradiated nLGADs is shown in \autoref{subfig:Isotherm_n} and \autoref{subfig:Isotherm_p}. For both samples, a slight initial reduction in depletion voltage is visible, corresponding to short-term beneficial annealing, which temporarily compensates for irradiation-induced damage before being overtaken by the reverse component. The annealing behavior observed in the irradiated nLGADs is consistent with the three-component model describing the change in effective doping concentration $\Delta {\mathrm{N_{eff}}}$ \cite{Moll1999}, separating the damage evolution into a short-term beneficial annealing component, a stable damage component, and a reverse annealing term.

\section{Conclusions}
\label{conclusion}
This work presents a systematic study of \SI{60}{MeV} proton and neutron irradiation effects on n-type Low Gain Avalanche Detectors (nLGADs), focusing on electrical characterization and annealing. The tested fluences range from \SI{5e12}{} up to \SI{1e14}{\text{particles}/\centi\meter\squared}. First I-V and C-V measurements before irradiation were performed, showing besides basic sensor properties, the temperature dependence of impact ionization and a clear shift in breakdown voltage with changing measurement temperature.

After irradiation, the I-V and C-V characteristics reveal changed sensor properties. The irradiation-induced changes can be explained by space charge sign inversion (SCSI) of the n-type silicon bulk. Measurements on both nLGADs and reference PiN diodes confirm that SCSI occurs within the investigated fluence range and highlight the unique mechanisms in n-type devices with a gain layer. SCSI alters the internal electric field configuration, leading to a reversed depletion direction and fundamentally different behavior compared to conventional p-type LGADs. C-V measurements reveal a distinct, fluence and frequency dependent peak near full depletion. This peak originates from the merging of two depletion regions forming at the front-side and back-side of the sensor and provides insight into the evolution of the internal field profile.

Post-irradiation annealing studies, conducted under both isochronal and isothermal conditions, show beneficial and reverse annealing. The evolution of the depletion voltage with annealing time matches well the established annealing models for silicon sensors.

Overall, the results deepen the understanding of nLGAD behavior under irradiation and provide guidance for their use in radiation environments. These findings can be relevant for detector development in HEP, but also for applications in areas such as space instrumentation and nuclear diagnostics. Beyond practical considerations, the study offers insight into the complex phenomenology emerging from the combination of n-type bulk material and a gain layer particularly under conditions of SCSI. 

\section*{Acknowledgments}
This work has received support from the European Union’s Horizon Europe Research and Innovation Programme under Grant Agreement No. 101057511 (EURO-LABS). We gratefully acknowledge IMB-CNM for providing the samples, which have received funding from from the Spanish Ministry of Science, Innovation and Universities (MICIU/AEI/10.13039/501100011033) through the grants PID2020-113705RB-C32, PID2023-148418NB-C42, and PDC2023-145925-C32.


\end{document}